# Anomalous superconducting properties at magic doping levels in under-doped $La_{2-x}Sr_xCuO_4$ single crystals

F. Zhou[a,*], P.H. Hor[b], X.L. Dong[b,a], W.X. Ti[a], J.W. Xiong[a] and Z.X. Zhao[a]

[a] *National Laboratory for Superconductivity, Institute of Physics,*
*Chinese Academy of Sciences, Beijing 100080, China*

[b] *Texas Center for Superconductivity and Department of Physics,*
*University of Houston, Houston, TX 77204-5002, USA*

**Abstract**

A series of high-quality under-doped $La_{2-x}Sr_xCuO_4$ superconductor crystals with x = 0.063 ~ 0.125 were prepared by traveling-solvent floating-zone (TSFZ) technique. We found by dc magnetic measurements that, in this series of crystals, the superconducting transition was quite sharp in the vicinity of the hole densities of x = 1/16 and x = 1/9 while it was much broader away from these two '*magic numbers*', and the Meissner fraction showed a remarkable minimum near x = 1/9. We concluded that these phenomena are reflections of intrinsic properties of this cuprate system. Our observations are discussed in light of recently proposed composite charge model together with charge inhomogeneity and electronic phase separations.

## 1. Introduction

Among the cuprate superconductor families, $La_{2-x}Sr_xCuO_4$ (LSCO-214) is known to be one of the systems that exhibit wide charge doping range while having fewer components and a simpler layered structure of $K_2NiF_4$ type with single $CuO_2$ planes. The overall density of doped holes in the $CuO_2$ planes of LSCO can be continuously tuned by substitution of $Sr^{2+}$ cations for $La^{3+}$ ones. Its electronic properties are strongly doping dependent [1-4] and the superconductivity occurs in the doping range of x = 0.06 ~ 0.26. While carrier doping dependences of physical properties in both superconducting and normal states are important for our understanding of high-$T_C$ superconductivity, observations of the charge inhomogeneity [5-7] and, especially, the intrinsic electronic phase separations [8, 9] imply that there are intrinsic instabilities of doped holes and that the free-carrier and superfluid densities can not be directly accounted for by a simple 'addition' of all the doped holes. In fact, most recently a composite charge picture [10] has been proposed that only a very small fraction of the total holes contributes to the free charge transport and superconductivity while the rest of the holes are in an ordered bound state forming a 2D electronic (charge) lattice. Different from the common CDW model, a key point of this composite charge picture is that the free carriers innately coexist with and move on the underlying 2D charge lattice that is formed at ~ 200K, which is much higher than $T_C$, in the $CuO_2$ planes [10]. Indeed, around room temperature, we have observed indications of hole localization at special hole concentrations of x = 0.06 (~1/16) and x = 0.1 (~1/9) by Hall effect measurements on oxygen-doped La-214 system [11]. A model based on the formation of 2D Wigner lattices

---

in the cuprates at x = 1/16 and x = 1/8, with $T_C$ = 15K and 30K respectively, was proposed from far-infrared charge dynamics study of strontium and oxygen co-doped La-214 [12, 13]. Under the framework of composite charge system, above observations thus suggest that the study on intrinsic electronic properties at these special hole-doping levels is extremely important for the ultimate understanding of the occurrences and mechanism of high $T_C$. Since all of the above observations were made in polycrystalline samples with a combination of hard- and soft-doping [8] and if they are truly intrinsic properties, we expect to observe experimentally more telling manifestations of our proposed physical picture by studying the electronic properties in a series of purely hard-doped and high-quality LSCO single crystals, especially in the close vicinity of these special charge densities. Among different techniques, the traveling-solvent floating-zone (TSFZ) method is accepted as a unique approach to growing such incongruent melting cuprate crystals. This crucible-free technique avoids the crystal contamination from crucible materials that will otherwise interfere with the observation of intrinsic electronic properties.

With this aim, we have made much effort and have succeeded in preparing by TSFZ technique a series of large and high-quality $La_{2-x}Sr_xCuO_4$ single crystals covering these special under-doping levels (x = 0.063 ~ 0.125). We report and discuss here our preliminary dc magnetic measurements that indeed reveal some unusual anomalies of superconducting properties in the vicinity of x = 1/16 (=0.0625) and x = 1/9 (=0.111) in this series of crystals. Discussions based on recent high-resolution ARPES data obtained on our crystal are also included.

## 2. Experimental

High purity (≥99.99%) oxides $La_2O_3$, CuO and carbonate $SrCO_3$ were used as the raw materials for the feed rods and the solvents used in crystal growth. For the feed rods, the initial composition was (1-x/2)$La_2O_3$/x$SrCO_3$/CuO in molar ratio with various dopant contents (x = 0.063, 0.07, 0.09, 0.10, 0.11, 0.125). Excess CuO of 1 ~ 2 mol% was added into the starting mixtures for compensating its evaporation loss in the growth process. The preparation of dense and homogeneous ceramic feed rods is one of the key factors in achieving a stable and continuous TSFZ growth. Much effort was made to establish the optimal preparation conditions. The starting powders were first well mixed and ground using a laboratory planetary ball mill and then prefired twice in air at 1000°C for 15 hours. After each thermal treatment, the materials were thoroughly ground again by ball milling. Thus obtained fine and homogeneous powders were put into silicon rubber tubes and hydrostatically pressed under a pressure > 600 MPa to form very compact cylindrical rods with a typical dimension of 6 ~ 7 mm in diameter and

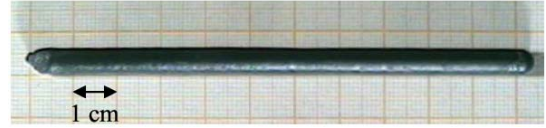

Fig.1. As-grown ingot of x ≈ 0.125 LSCO crystal.

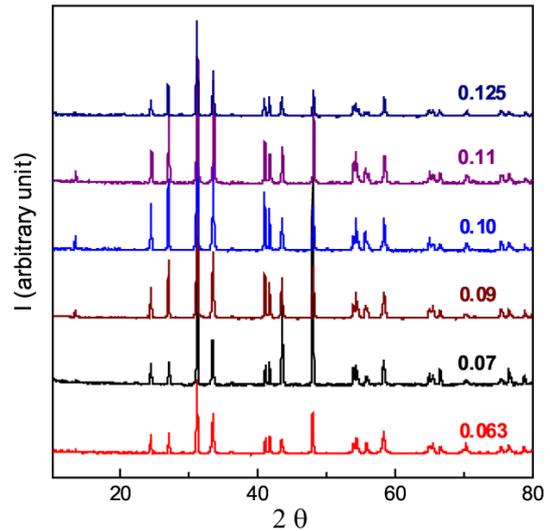

Fig.2. Powder XRD patterns for this series of $La_{2-x}Sr_xCuO_4$ single crystals.

135 mm in length. The pressed rods were finally sintered at 1210 ~ 1260°C for several hours in flowing oxygen using a specially designed vertical furnace (Crystal Systems Inc., VF-1800/EF-6000). For the solvents, the composition of staring materials was much richer in CuO as self-flux, which was typically of 78 mol% in content. In determining the Sr contents of the solvents corresponding to the various Sr doping levels of present interest, we took into account the previously reported results of distribution coefficients $k_{Sr}$ of Sr doping into $La_2CuO_4$ [14]. After ground and mixed the starting mixtures were prefired in air at 970°C for 15 hours, they were then ground and mixed again before pressed into pellets of about 700 mg. The final sintering of the solvent pellets was performed in air at 1000 ~ 1010°C for 12 hours.

An infrared-heating floating-zone furnace with a quartet ellipsoidal mirror (Crystal Systems Inc., FZ-T-10000-H) was used for TSFZ experiments. The crystals were grown under an oxygen pressure of 0.2 MPa at a zone traveling rate of 0.8 mm/hr using seed crystals orientated along orthorhombic or tetragonal [100] or [110] directions. During the growth the feed rods and the grown crystals were continuously rotating in opposite directions at 20 ~ 30 rpm to improve the heating uniformity and melt

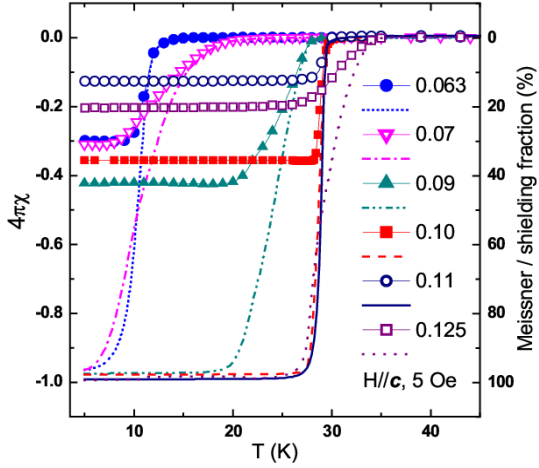

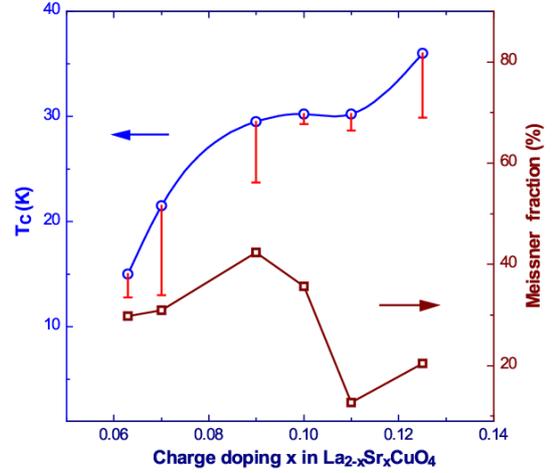

Fig.3. Meissner curves (symboled lines) and shielding curves (bare lines) of $La_{2-x}Sr_xCuO_4$ single crystals with various under-doping levels (x = 0.063 ~ 0.125).

Fig.4. The onset superconducting transition temperature $T_C$, the transition width (indicated by the length of the vertical bars in the $T_C$ curve) and the Meissner fraction as a function of the hole concentration x for $La_{2-x}Sr_xCuO_4$ crystals. The data were derived from the dc magnetic measurements shown in Fig 3. Lines are guide to the eye.

homogeneity.

The dc magnetic measurements were performed on SQUID magnetometer (Quantum Design, MPMS-XL) with **c**-axis of the crystals along the magnetic field. Powder XRD patterns were registered using a diffractometer equipped with an 18 KW rotating Cu target (Mac Science, M18AHF). The crystal compositions were checked by the inductively coupled plasma atomic emission spectroscopy (ICP-AES). Optical microscopic observations were done on a microscope with a polarized lighting system.

## 3. Results and Discussion

By careful operations and the use of high quality ceramic feed rods, a stable molten zone was successfully maintained until the end of the growth in our most TSFZ experiments. Shown in Fig.1 is the photo for one of the as-grown ingots, with a typical size of 5 ~ 6 mm in diameter and 110 mm in length. Crystal pieces cut from the ingots and polished along (100), (110) and (001) crystal planes were carefully checked by optical microscopy using polarized and normal light, revealing that large single-grain crystals were obtained. The compositions of the grown crystals estimated by ICP-AES were quite close to those of the feed rods, and no foreign phases in crystal samples were detected by powder XRD analysis, as can be seen from Fig. 2.

The Meissner (field-cooled) and the shielding (zero-field-cooled) signals of all the six crystal samples were measured in a low field of 5 Oe on warming and are shown in Fig. 3. The magnetic data were corrected for demagnetizing factors. Given in Fig. 4 are the onset superconducting transition temperature, the transition width (defined here as the difference between the data at 10% and 90% of full Meissner signal) and the Meissner fraction as a function of the hole concentration x. The critical temperature, $T_C$, shows a familiar evolution with x but a 'plateau' appears at around x = 1/9. It is of much interest to note that the superconducting transition width as well as volume fraction vary with x by no means monotonically but anomalously. Near the doping levels of x = 1/16 (=0.0625) and x = 1/9 (=0.111), the transitions are much sharper ($\Delta T \approx 2$ K) than those away from these two '*magic numbers*' ($\Delta T \geq 6$ K), and the Meissner fraction drops remarkably, accompanying the $T_C$ plateau, to a minimum in the vicinity of x = 1/9. The shielding signals exhibit the same broadening trend for the doping levels away from the magic numbers and they all have 100% volume fraction. This indicates that there are no macroscopic inhomogeneity and/or weak links.

It needs to be pointed out that, based on our careful examinations on the representative x = 0.09 crystal showing a broad SC transition, we ruled out the possibility that such transition broadening may result from crystalline imperfection [15]. We therefore concluded that the anomalous doping dependences in this series of crystals are intrinsic properties of the system. Furthermore, recent high-resolution ARPES experiments along the nodal direction [16] have seen, even above $T_C$, in our x ≈ 1/16 crystal a clear drop at ~70 meV in the quasiparticle scattering rate [Fig. 1b of ref. 16], which is necessarily related to a counterpart kink in the dispersion [17], and have revealed that the scattering rate is relatively low, an indication of high electronic quality of the sample. This is true for our other crystals in the present doping series. We now discuss our results in terms of our previous observations in the co-doped systems. First, we note that the superconducting transition temperatures are 15K and 30 K for the two special doping levels near $x = 1/16$ and $x = 1/9$, respectively (see Fig. 4). They are exactly the two intrinsic $T_C$'s observed earlier [8]. In other words, these two superconducting phases have been preferentially formed at around the magic doping levels. That can naturally explain the single sharp SC transitions near the two magic doping levels and the broadening of the transition while away from these magic dopings. Indeed, the 15K SC transition in our crystal at $x \approx 1/16$ is the same $T_C$ that we observed in our cation and anion co-doped system at exactly the same carrier concentration where the formation of a p(4x4) 2D Wigner lattice was proposed [12, 13]. For $x \approx 0.125$ crystal, the transition is not sharp and the Meissner signal is small. This is probably due to the competition of the stripe phase formation near this concentration. Indeed, we have observed a clear competition between stripe phase superconductivity and 30 K transition in a series of well-prepared stripe phase samples [18]. However, while the sharp single superconducting transitions are easily understood in terms of our model, the anomalous drop of Meissner signal at $x = 1/9$ seems to be very unusual. Further work is underway.

## 4. Conclusion

In a series of high-quality under-doped $La_{2-x}Sr_xCuO_4$ crystals, we have observed from the dc magnetic measurements that the superconducting transition is quite sharp in the vicinity of two magic hole-doping levels of $x = 1/16$ and $x = 1/9$ while remains broad at the other hole densities, and the Meissner signal fraction exhibits an unusual decrease near $x = 1/9$. We conclude that these interesting anomalous behaviors are intrinsic properties of this cuprate system. Our observations are discussed in light of our reported intrinsic electronic superconducting phase separations [8] and recently proposed composite charge model [10] including naturally charge inhomogeneity.

## Acknowledgments

We would like to thank Prof. Z.L. Wang and Mrs. N.L. Li for various and valuable helps in crystal orientation and processing, and Mrs. H. Chen for XRD experiments. The work in Beijing is supported by *Ministry of Science and Technology of China* (Project G1999064601) and *National Natural Science Foundation of China* (Project 10174090). The work in Houston is supported by *the State of Texas* through *The Texas Center for Superconductivity*.